\documentclass[conference]{IEEEtran}
\IEEEoverridecommandlockouts
\usepackage{cite}
\usepackage{amsmath,amssymb,amsfonts}
\usepackage[ruled,linesnumbered,vlined]{algorithm2e}
\usepackage{graphicx}
\usepackage{textcomp}
\usepackage{xcolor}
\usepackage{listings}
\usepackage{graphicx}
\usepackage[most]{tcolorbox}
\usepackage{mathtools}
\usepackage{subfigure}
\usepackage{array}
\usepackage{tikz}
\usepackage{pifont}
\usepackage{bm}
\usepackage[T1]{fontenc}
\usepackage{soul}
\usepackage[normalem]{ulem}
\usepackage{blkarray}
\usepackage{multirow}
\usepackage{color, colortbl}
\usepackage{url}

\SetKwRepeat{Do}{do}{while}

\SetCommentSty{mycommfont}

\definecolor{Gray}{gray}{0.9}

\newcommand*\circled[1]{\tikz[baseline=(char.base)]{
            \node[shape=circle,draw, inner sep=0.7pt,  anchor=center] (char) {#1};}}

\newcolumntype{P}[1]{>{\centering\arraybackslash}p{#1}}
\newcolumntype{M}[1]{>{\centering\arraybackslash}m{#1}}

\newtcbox{\dashedbox}[1][]{
  baseline=0.4\baselineskip,
  equal height group=dashedbox,
  nobeforeafter,
  colback=white,
  boxrule=0pt,
  enhanced jigsaw,
  boxsep=0pt,
  top=1pt,
  bottom=1pt,
  left=3pt,
  right=3pt,
  borderline horizontal={0.5pt}{0pt}{dashed},
  borderline vertical={0.5pt}{0pt}{dashed},
  #1
}

\definecolor{codegreen}{rgb}{0,0.6,0}
\definecolor{codegray}{rgb}{0.5,0.5,0.5}
\definecolor{codepurple}{rgb}{0.58,0,0.82}
\definecolor{backcolour}{rgb}{0.95,0.95,0.92}
 
\lstdefinestyle{mystyle}{
    backgroundcolor=\color{backcolour},   
    commentstyle=\color{codegreen},
    keywordstyle=\color{magenta},
    numberstyle=\tiny\color{codegray},
    stringstyle=\color{codepurple},
    basicstyle=\ttfamily\footnotesize,
    breakatwhitespace=false,         
    breaklines=true,                 
    captionpos=b,                    
    keepspaces=true,                 
    numbers=left,                    
    numbersep=5pt,                  
    showspaces=false,                
    showstringspaces=false,
    showtabs=false,                  
    tabsize=2
}
 
\def\BibTeX{{\rm B\kern-.05em{\sc i\kern-.025em b}\kern-.08em
    T\kern-.1667em\lower.7ex\hbox{E}\kern-.125emX}}
    
\begin{document}
\setstcolor{red} 

\newcommand{\pname}{\textsc{IoTGaze }}

\title{\textsc{IoTGaze}: IoT Security Enforcement via Wireless Context Analysis
}
\author{
    \IEEEauthorblockN{
        Tianbo Gu\IEEEauthorrefmark{1},
        Zheng Fang\IEEEauthorrefmark{1},
        Allaukik Abhishek\IEEEauthorrefmark{2},
        Hao Fu\IEEEauthorrefmark{1}, 
        Pengfei Hu\IEEEauthorrefmark{1},
        Prasant Mohapatra\IEEEauthorrefmark{1}
    }
    \IEEEauthorblockA{
        \IEEEauthorrefmark{1}
        Department of Computer Science, University of California, Davis, CA, USA\\
        \IEEEauthorrefmark{2}
        ARM Research, Austin, TX, USA\\
        Email: \{tbgu, zkfang, haofu, pfhu, pmohapatra\}@ucdavis.edu, Allaukik.Abhishek@arm.com
    }
}
\maketitle

\begin{abstract}
Internet of Things (IoT) has become the most promising technology for service automation, monitoring, and interconnection, etc. However, the security and privacy issues caused by IoT arouse concerns. Recent research focuses on addressing security issues by looking inside platform and apps. In this work, we creatively change the angle to consider security problems from a wireless context perspective. We propose a novel framework called \pname, which can discover potential anomalies and vulnerabilities in the IoT system via wireless traffic analysis. By sniffing the encrypted wireless traffic, \pname can automatically identify the sequential interaction of events between apps and devices. We discover the temporal event dependencies and generate the \textit{Wireless Context} for the IoT system. Meanwhile, we extract the \textit{IoT Context}, which reflects user's expectation, from IoT apps' descriptions and user interfaces. If the wireless context does not match the expected IoT context, \pname reports an anomaly. Furthermore, \pname can discover the vulnerabilities caused by the inter-app interaction via hidden channels, such as temperature and illuminance. We provide a proof-of-concept implementation and evaluation of our framework on the Samsung SmartThings platform. The evaluation shows that \pname can effectively discover anomalies and vulnerabilities, thereby greatly enhancing the security of IoT systems.

\end{abstract}

\begin{IEEEkeywords}
Internet of Things, Anomaly Detection, IoT Security, Natural Language Processing, Wireless Context.
\end{IEEEkeywords}

\section{Introduction}
The rapid development of the Internet of Things (IoT) has an increasingly bigger impact on how we live and work. IoT technology enables interconnection, service automation, and other convenience in a variety of application scenarios, such as smart home, smart factory, and smart city, etc. By 2022, the number of connected IoT devices will reach to 29 billion \cite{erric2018}. The market value of IoT will reach to $\$1.2$ trillion in 2022 with a compound annual growth rate of 13.6\% starting from 2017 according to the IDC prediction \cite{idc2018}. To increase their market share, different companies develop their IoT platforms for third-party developers to build apps to realize service provision automation. The popular IoT program platforms include Samsung's SmartThings \cite{smart2015}, Apples' HomeKit \cite{apple2015} and Google Home \cite{google2015}.

Despite the exploding devices and fast growth of platforms of IoT, the security and privacy solution is not keeping the pace. Emerging vulnerabilities and attacks in IoT have brought tremendous loss. Within 20 hours, 65,000 IoT devices were rapidly infected and utilized to launch Mira attacks leading to internet outage \cite{antonakakis2017understanding}. By exploiting a major bug in the implementation of the \textit{ZigBee} light link protocol, the attacker can use one single malicious bulb to turn off all the city lights \cite{ronen2017iot}. Most critical security and privacy threats come from the IoT platforms and their affiliated apps.  For instance, despite the Samsung SmartThings platform has a capability separation model, the apps can still request the capabilities that they do not need. The platform lacks effective means to audit the requests.  The authors \cite{fernandes2016security} found that 55\% of SmartApp did not use all the rights to device operations that their requested capabilities implied, and 42\% of SmartApps were granted capabilities that were not explicitly requested or used. Once gaining access to the capabilities, the malicious apps may not follow the user expectation and their app descriptions, resulting in serious security issues.

To relieve the security and privacy threats, the researchers propose solutions from different perspectives. By embedding extra code, FlowFence \cite{fernandes2016flowfence} and Soteria \cite{celik2018sensitive} can monitor the data flows and related control flows to prevent all the implicit flows from IoT apps via static program analysis. ContextIoT \cite{jia2017contexlot} uses the runtime logging to extract the essential context for building a context-based permission system. SmarthAuth \cite{tian2017smartauth} collects the security-relevant context information from analyzing IoT apps' source code, annotations, and descriptions.
IoTGuard \cite{celik2019iotguard} dynamically collects the apps' information to enforce safety and security policies. However, these approaches require good knowledge about the program framework and app code. They have to modify the apps' source code or patch the apps and platforms to realize the discovery and prevention of threats. As can be seen, most existing solutions focus on the program analysis for platforms and apps. Then we come up with a question: \textit{Can we open a new path to enhance the defense of IoT security and privacy?}

In this work, we look outside the IoT platforms and apps, and rethink the IoT security and privacy problems from the wireless perspective, and propose a new concept \texttt{Wireless Context} in IoT. Distinct from the program-based context, the IoT \textit{Wireless Context} is inferred from the wireless communication traffic. We propose and implement a novel IoT security enforcement framework called \pname that can detect potential anomalies and vulnerabilities in the IoT system. First, \pname extracts the wireless packet features to correlate the communication traffic with the interaction of events between apps and devices. \pname constantly sniffs the encrypted wireless traffic and generates the interaction event sequence. Second, \pname discovers the temporal event dependencies and builds the \textit{Wireless Context} for IoT system. Third, \pname extracts the actual user expected \textit{IoT Context} from IoT apps' descriptions and user interfaces (UI). By comparing the detected \textit{Wireless Context} with \textit{IoT Context}, \pname can discover the anomaly in current IoT system. Lastly, by exploring the wireless event dependencies, \pname is able to discover the unknown vulnerabilities that are caused by the inter-app interaction chain via hidden channels, such as light, temperature, humidity, etc., and can be exploited by the attacker to launch attacks against IoT system.

\textbf{Contributions:} The contributions of our work are:

\begin{itemize}
    \item Distinct from the existing solutions, we open a new path to rethink the IoT platform and app security issues from the \texttt{Wireless Context} perspective and propose a novel IoT anomaly and vulnerability detection framework called \pname.
    \item We design a fingerprinting approach to detect the IoT events and generate the sequence of the events via analyzing the wireless packets. We also propose an effective mechanism to discover the temporal events dependencies and produce the event transition graph that represents the \textit{Wireless Context} in IoT.
    \item We propose an approach that can extract the user expected \texttt{IoT Context} from apps' descriptions and UI using natural language processing (NLP). An algorithm is designed to detect the anomaly based on the comparison between the \textit{IoT Context} and \textit{Wireless Context}.
    \item By exploring \textit{Wireless Context}, \pname can discover the hidden vulnerabilities that are caused by the inter-app interaction, which is ignored by most exiting IoT security solutions.
    \item We prototype a proof-of-concept framework on the Samsung SmartThing platform, including 183 apps. The extensive evaluations show that our approach can achieve nearly 98$\%$ accuracy of anomaly detection. We also discover and provide a complete list of hidden vulnerabilities in the IoT system detected by \pname.
\end{itemize}

\section{Threat Model}
In this paper, we consider the security and privacy problems on the typical IoT interaction chain: devices, apps, and IoT platform. Based on the program framework, the developers write apps that request the capabilities access privilege from devices, and then control the devices to implement service automation. For instance, the description of one app is \textit{``Turn on the indoor surveillance if the householder leaves home, otherwise turn off the indoor surveillance''}. The related IoT devices are presence sensor and surveillance camera. The security and privacy issues for the IoT system we want to detect are: \textbf{(a)} \texttt{App misbehavior}. For instance, when the household is at home, the app should turn off the surveillance to prevent privacy leakage. But the app may not turn off the surveillance and still monitor the activities of the household and uploads the data to somewhere else. \textbf{(b)} \texttt{Event spoofing}. The attacker may spoof a \textit{``presence.not\_present"} command to the IoT hub and the hub turns off the surveillance. Then the intruder could break into the house. \textbf{(c)} \texttt{Over-privilege}. The app may request irrelevant capabilities from the platform, such as the lock control privilege. Then the app may unlock the door when the household leaves home, which triggers serious security issues. \textbf{(d)} \texttt{Device failure}. The hardware flaws and software bugs may cause device failure. The attacker can also launch an attack to make the device (e.g., surveillance camera) unresponsive. \textbf{(e)} \texttt{Hidden vulnerabilities}. This type of vulnerabilities is caused by some hidden channels that multiple apps interact with simultaneously. Consider the scenario where one heater control app can automatically turn on the heater in winter after 8:00 PM, and another app opens the window automatically if the room temperature is higher than 90$^\circ$. The indoor temperature is the physical channel, and the attacker can spoof a command event to let the heater keep working, leading to an increase of the temperature, and finally open the window and break in. We design a novel anomaly and vulnerability detection framework called \pname that addresses the above security and privacy threats in IoT system.
\section{System overview}
\begin{figure}[t]
\centering
\includegraphics[scale=0.19]{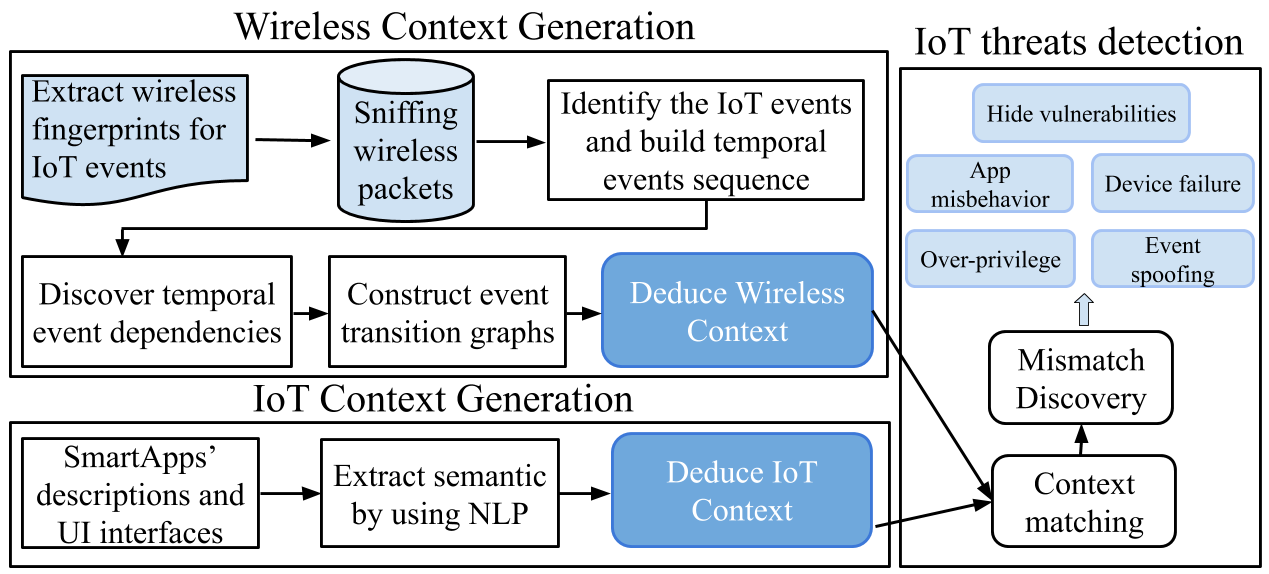}
\caption{Overview of \pname}
\label{fig:arch}
\end{figure}

In this section, we provide an overview of \pname, and describe key components and workflow, as shown in Fig. \ref{fig:arch}. 

\textbf{Wireless Context Generation}. The challenge here is how to use the sniffed raw wireless packets to generate the IoT \texttt{Wireless Context}. We decompose this problem into three subtasks: (1) \textit{How to correlate the wireless traffic with the IoT interaction events and generate the fingerprints for the events?} We utilize limited features extracted from the encrypted wireless traffic and generate effective fingerprints to detect IoT events. (2) \textit{How to use the sequential packets to generate the sequential IoT events?} What we sniff is the wireless packet sequence, but for vulnerability detection we should work on events. Thus, we design an approach to automatically segment the packet sequence and generate the temporal event sequence. (3) \textit{How to discover the temporal event dependencies and generate the wireless context?} We design an event dependency discovery method that accurately extracts the event dependencies and their causal relationship for building the wireless context graph.

\textbf{IoT Context Generation}. The \texttt{IoT Context} we define here is the event interaction chain between smart apps (which run in the cloud and interact with devices via the IoT gateway such as SmartThings Hub) and devices. The IoT context is the user expected app behaviors, which may not be the real execution behaviors of apps. The malicious apps may deceitfully inform users about their functionalities but surreptitiously execute some malicious activities. To accurately extract the IoT context, we analyze the apps' descriptions and UIs that are directly exposed to users and usually cannot deceit users compared with program code. We extract the IoT context from app description and UI using NLP techniques and build the corresponding event transition graph that represents the app work logic expected by the user.

\textbf{Anomaly and Vulnerability Detection}. The IoT context represents the IoT automation services expected by the user, while the wireless context reveals what practical automation services are happening. Each context is expressed by a set of event transition graphs. We propose an approach to discover the mismatch and anomaly by comparing the event transition graphs under different contexts. By further analysis, we can discover the hidden vulnerabilities that are caused by the inter-app interaction and can be used by an attacker to launch attacks. Then we can prevent the attacks before they happen. Next, we describe these components of \pname in detail in the following section.
\section{Wireless Context Generation}

We design and deploy a third-party \textit{guardian} who \textit{gazes} at the wireless communication traffic and detect potential anomalies and vulnerabilities. The \textit{guardian} sniffs the encrypted wireless packets generated by the IoT activities and record the packets sequence $P=\{p_1, p_2, ...,p_i, ..., p_n\}$. Our goal here is to analyze the packet sequence and generate the wireless context. The wireless context is represented by a set of event transition graphs. In this section, we explain the procedures of wireless context generation in detail.

\subsection{IoT Event Fingerprinting}\label{sec:fingerprint}
Before generating the event sequence, we need to correlate the wireless traffic with the IoT events. We design the fingerprints to identify the IoT events. To realize service automation, event-driven smart apps receive data from various sensors (such as motion sensor, temperature sensor, and contact sensor) and issue commands to one or more actuators (e.g., smart bulb, smart power outlet, and smart lock, etc.) via the local IoT hub as the intermediary. We define IoT events as the activities that IoT hub interacts with sensors and actuators via wireless communication.

We use a packet sequence $P_{e_i} = \{p_1,p_2,...,p_i,...,p_{N_{e_i}}\}$ to represent the sniffed traffic for a unique event $e_i$. We extract the features from the packets' attributes except the encrypted data content to fingerprint the event. We list the features as following:(1) \textit{Packet size}. A packet size could vary depending on what it transmits for which event. (2) \textit{Packet direction}. A packet could be sent from the hub to a device or the opposite way. (3) \textit{Packet interval}. The shorter packet interval signifies the higher transmission rate. Due to the difference of software and hardware, IoT devices may have distinct transmission rate, burst rate, response latency, and throughput, leading to varying packet interval. (4) \textit{Packet layer}.  Packets may be transmitted in different layers for a specific protocol. The above are common features across various wireless communication protocols, such as WiFi, Zigbee, Z-Wave, and Bluetooth lower Energy (BLE). Each protocol may have additional features. For example, The IP-based communication protocol may have features like IP source/destination address and source/destination port. By using the features set, we can generate the following fingerprint for a unique IoT event:
\vspace{-2mm}
\[
F_{e_{i}} = 
\begin{blockarray}{ccccc}
p_1 & \cdots & \cdots & p_{N_{e_i}} \\
\begin{block}{(cccc)c}
f_{1,1}&f_{2,1} & \cdots  &f_{N_{e_i},1} \\
f_{1,2}&f_{2,2} & \cdots  &f_{N_{e_i}, 2} \\
\vdots  & \vdots & \ddots & \vdots\\
f_{1,\mathcal{M}}&f_{2,\mathcal{M}} & \cdots  &f_{N_{e_i},\mathcal{M}} \\
\end{block}
\end{blockarray}
\]
\vspace{-6mm}

\noindent where $N_{e_i}$ denotes the number of packets transmitted for event $e_i$, and $\mathcal{M}$ denotes the number of features we extract for a specific communication protocol.

We collect and create the fingerprints data set for each event, and use the Random Forest supervised machine learning model as the classifier $\mathcal{C}$. The value of $N_{e_i}$ varies depending on the event $e_i$. In order to feed the fingerprints matrix $F_{e_i}$ into the same machine learning model, we fix the number of packets to $\mathcal{N}$ and pad the matrix with zero if $N_{e_i}$ is less than $\mathcal{N}$. The optimal value of $\mathcal{N}$ will be discussed and selected in the later evaluation section. Then we use the classifier $\mathcal{C}$ to classify the new, unlabeled fingerprints and identify the events.

\begin{figure}[t]
\centering
\includegraphics[scale=0.24]{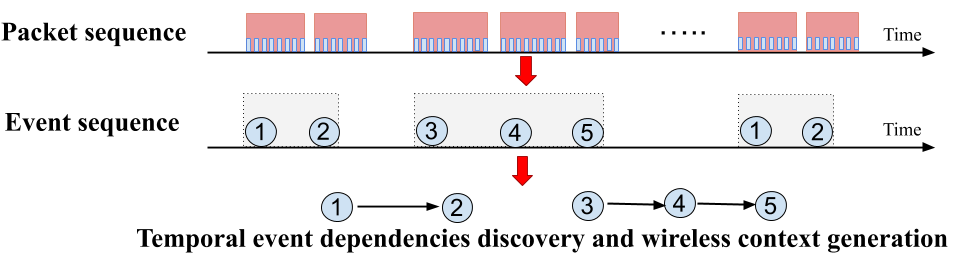}
\caption{Procedures of wireless context generation.}
\label{fig:wirelesscontext}
\end{figure}

\subsection{Sequential Events Generation}\label{sec:event_identifcation}
We analyze the sniffed packets sequence:
\begin{equation}
P = \{(p_1, t_1),(p_2, t_2),...,(p_i, t_i),...,(p_n, t_n)\},
\end{equation}
and identify the events using the generated fingerprints. Considering the packets for an event are sent within a short time. We use a sliding window with a maximum $\mathcal{N}$ packets within a fixed time interval of $\mathcal{T}$, and produce the matrix $F$. Then we feed the matrix $F$ to the classifier $\mathcal{C}$ and output the probabilities for each event. If the maximum probability value predicted from event $e_j$ is larger than the predefined classification threshold $\theta$, then we think the packet sequence is created by event $e_j$. Otherwise, we will go to the next sliding window and continue to make the identification. The default step size is one, and the step size changes to the number of packets from event $e_j$ once event $e_j$ is detected.

\subsection{Temporal Event Dependencies Detection}
After identifying individual wireless events, we can construct the event stream $E = \langle(e_1,t_1),(e_2, t_2),...,(e_n,t_n)\rangle$, where $e_i$ ($i = 1, \dots, n$) is the wireless event happening at time $t_i$. Notice that $e_i$ and $e_j$ ($i\neq j$) can be the same event happening at different time. A temporal event dependency means a set of events occur together with a chronological pattern. If event type $a$ and $b$ have a temporal dependency, then the time interval between them should follow a normal distribution $\mathcal{N}(\mu(a,b),\,\sigma^{2})$ with $\sigma$ being approximately equal to the standard deviation of the network delay. Thus, even though the expectation $\mu$ depends on the particular event type, the standard deviation is independent of event types. To determine if $a$ and $b$ are temporally dependent, we can collect all the samples of time interval between $a$ and $b$ from the event stream, and compute the sample standard deviation $\sigma(a,b)$ and compare it with the threshold $\tau$ ($\tau$ is a predefined parameter which is slightly larger than the standard deviation of network delay). If $\sigma(a,b)<\tau$, then we conclude $a$ and $b$ are temporally dependent, and vice versa.

Once we have identified all the pairs of events that are temporally dependent, we reconstruct the dependency sequence by concatenating these pairs. Formally, if $[a,b]$ and $[b, c]$ are dependent pairs, and $\mu(a,c) = \mu(a,b)+\mu(b,c)$, then we can get a dependency sequence $[a, b, c]$. Following such procedure, we iteratively check and concatenate sequences. In addition, we find that even if there is a dependency sequence $[a, b, c, d]$, $[c, d]$ itself could be a dependency sequence. We further discover these ``subsequences of dependency sequences" using the number of occurrence in the input event stream. For example, if $[a, b, c, d]$ occurs 100 times, $[b, c, d]$ occurs 100 times, but $[c, d]$ occurs 150 times, then we know that $[b, c, d]$ is not a dependency sequence (since it is just a part of the dependency sequence $[a,b,c,d]$), but $[c, d]$ is a dependency sequence by itself and it occurs 50 times.

\textbf{Generating Wireless Context}. After discovering the event dependencies, we can build the event transition graph for each event dependency. As shown in Fig. \ref{fig:wirelesscontext}, the event transition graph $\circled{\small 1}\xrightarrow{}\circled{\small 2}$ represents a certain wireless context, such as \textit{``If detecting a human presence, open the surveillance camera.''}. The wireless context is extracted from the wireless traffic and can reflect the real activities of the currently installed apps. However, the wireless may not the expected IoT context from the user. We introduce the approach to extract the IoT context in the following section. If the wireless context violates the IoT context expected by the user, then it indicates potential anomalies in the current IoT system.
\section{IoT context generation}\label{sec:context}
\begin{figure}[t]
\centering
\includegraphics[scale=0.1]{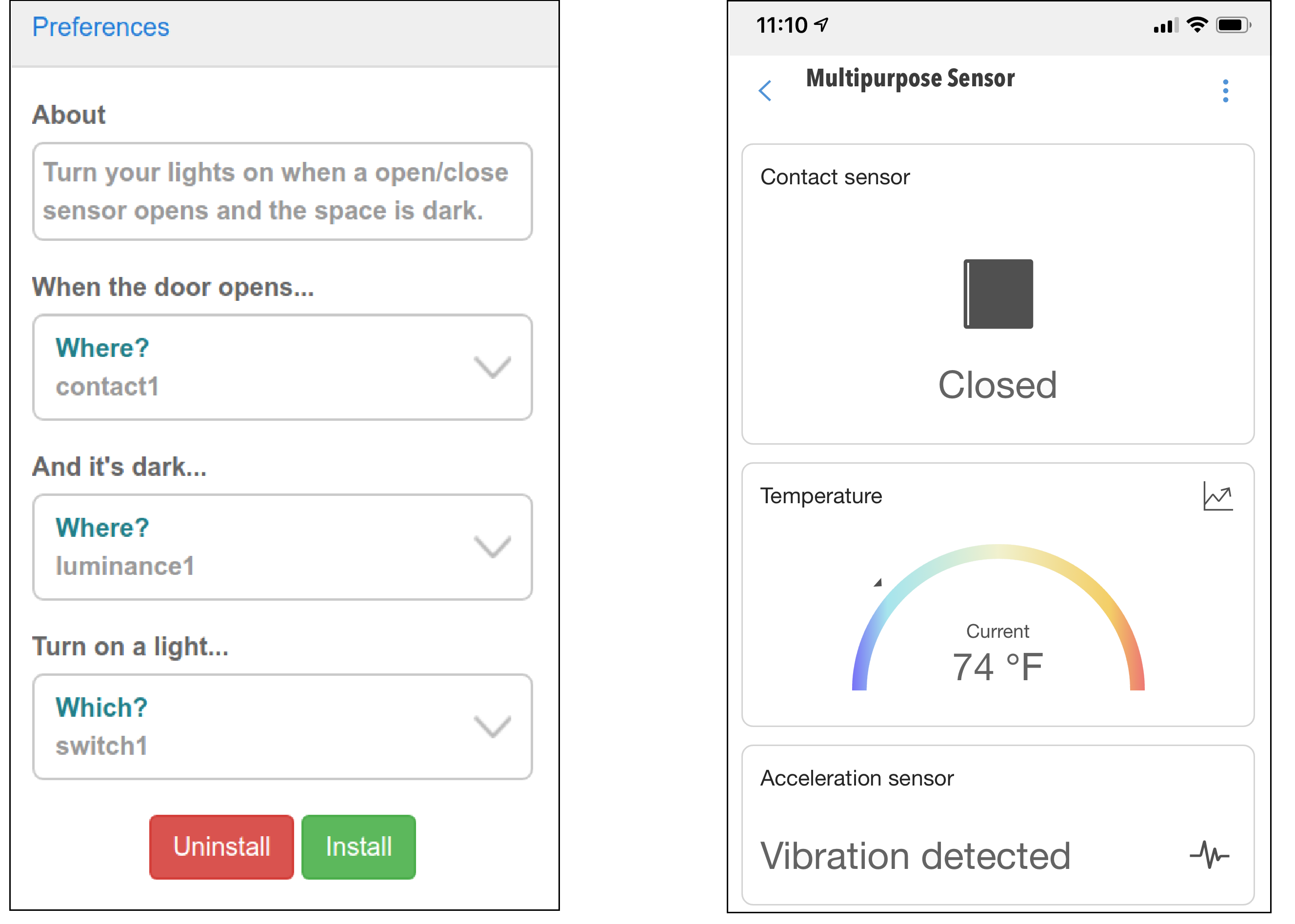}
\caption{(a) The installation interface of the SmartApp \textit{Brighten-Dark-Places} in Samsung SmartThings platform. (b) The capabilities that one multipurpose sensor has for the Samsung SmartThings platform.}
\label{fig:UI}
\end{figure}
In this section, we explain how to collect the IoT context that is the expected automation services from users. Due to the existing of malicious apps, the activities of smart apps cannot represent the actual IoT context. Some existing work conducts the static and dynamical analysis of the apps' code and checks if the actions of the program match what the apps describe. Instead of analyzing the apps' source code, we exploit the apps' description and UI that the apps usually do not tamper or spoof. Fig. \ref{fig:UI}(a) shows the installation interface of one SmartApp \textit{Brighten-Dark-Places}. Based on the app description, the user chooses to install the app or not. Meanwhile, the needed capabilities are requested by the app, and the user needs to select the devices that can provide the capabilities. As we can see, the content in the installation interface is directly exposed to the user and tells the user what the app plans to do and is not usually tampered. If the user chooses to install the app, that means the app's description can reflect the user truly expected app service. If we know all the smart apps installed by the user, we can build the IoT context based on these apps' descriptions and UI. Now, we introduce the IoT context generation approach and implement it on Samsung SmartThings platform \cite{smart2015}.


\subsection{App Description Analysis}

The research work \cite{nandi2016automatic,bastys2018if} has revealed that most IoT applications following the ``If-This-Then-That" (IFTTT) programming paradigm, which can also be reflected by their apps' description. The first step for analyzing apps' behaviors is to obtain the causal relationship from the description. One effective method is to identify the conditional and main clause from the description. The conditional clause involves some sensors' state change (e.g., the camera recognizes someone's face), and the main clause involves some devices' actions (e.g., unlock the door). Then, we can extract the related devices and their actions from the noun phrase and the verb phrase, respectively.

\begin{figure}[t]
\centering
\includegraphics[scale=0.22]{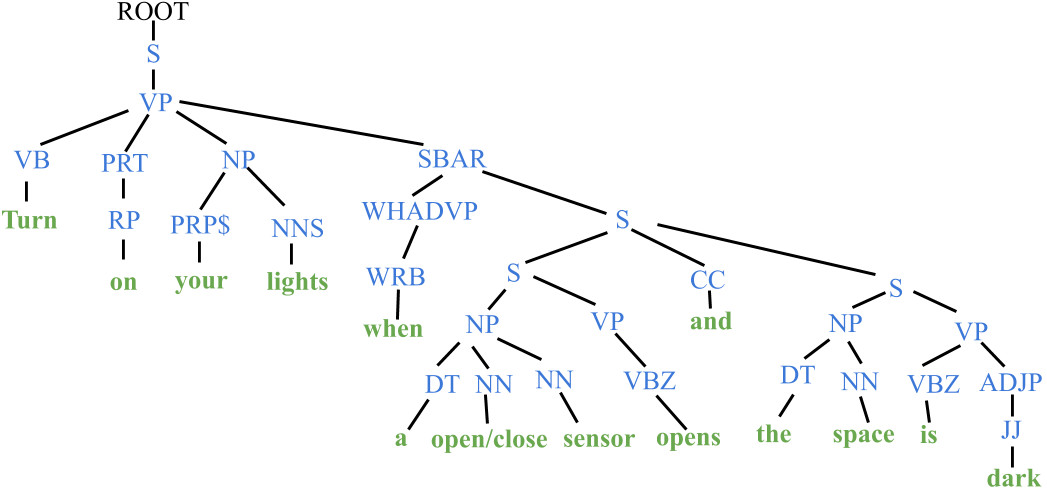}
\caption{Stanford constituency tree representation of the description from the Brighten-Dark-Places SmartApp.}
\label{fig:constituency}
\end{figure}

\begin{figure}[t]
\centering
\includegraphics[scale=0.2]{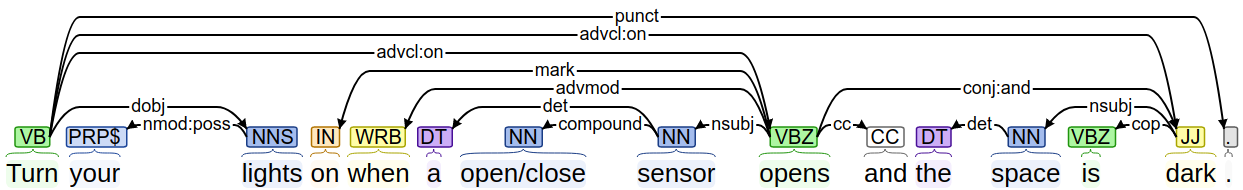}
\caption{Stanford dependency tree representation of the description from the Brighten-Dark-Places SmartApp.}
\label{fig:dependency}
\end{figure}

We use Stanford parser \cite{manning2014stanford} to analyze the sentence structure of the app descriptions. To segment the description sentence into clauses, we build the constituency parse tree and split the sentence by label \textbf{S} (\textit{Simple declarative clause}) or \textbf{SBAR} (\textit{Clause introduced by a subordinating conjunction}). As shown in Fig. \ref{fig:constituency}, the extracted three clauses are: ``Turn on your lights", ``a open/close sensor opens", and ``the space is dark". The subordinating conjunction ``when" signifies the causal relationship of the clauses via identifying the trigger and action. We analyze the dependency parse tree in Fig. \ref{fig:dependency} and extract the noun phrase and verb phrase from each clause. Considering the first clause as an example, ``lights" is the accusative object of the verb ``Turn" and this dependency is represented as \emph{dobj}(``Turn", ``lights"). But the extracted semantic may be human-readable and not machine-readable. We continue to do the capability matching process.


\subsection{Capability Matching}

The SmartApps interact with devices based on their capabilities. The capabilities have to be well decomposed in order to prevent over-privilege. The Samsung SmartThings platform maintains a capability list \cite{capasmart2018} that SmartApps can request. Fig. \ref{fig:UI}(b) shows the multipurpose sensor has three capabilities that can provide to apps: \emph{capability.contactSensor}, \emph{capability.temperatureMeasurement}, and \emph{capability.accelerationSensor}.
Although we have extracted the app behavior from the description, there could still be a semantic gap between the wording of the description and the capabilities. Hence, we need to establish the relationship between the non-phrases in the description and the capabilities. The verb phrase in the same clause may also provide useful information for the matching and could also be considered during the matching. For example, ``is dark" is more related to ``illuminance" than ``the space".




We match noun phrases and verb phrases to the capabilities based on the similarity score computed by the Word2Vec \cite{mikolov2013efficient} model trained on the part of Google News dataset (about 100 billion words). Because the Word2Vec only gives embedding for words, we split every phrase into a tuple of individual words. This operation is also performed for capability names. We take the highest score of all the possible word pairs between a phrase tuple and a capability tuple as the similarity of these two tuples. Once we have the similarity score for each phrase and capability pair, we match the clause to the most similar capability. For each clause, if the most similar capability is already taken by some other capabilities, the second most similar one is chosen. Taking \textsc{Brighten-Dark-Places} SmartApp as an example, the matching result is ``Turn on your lights" $\leftrightarrow$ \emph{capability.switch}, ``a open/close sensor opens" $\leftrightarrow$ \emph{capability.contactSensor}, and ``the space is dark" $\leftrightarrow$ \emph{capability.illuminanceMeasurement}.

\subsection{Event Transition Graph Generation}

\begin{figure}[t]
\centering
\includegraphics[scale=0.3]{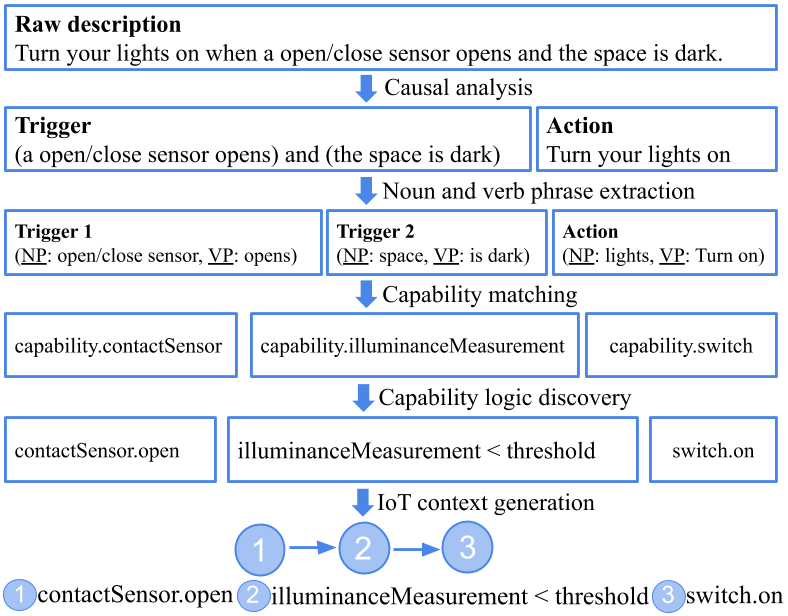}
\caption{IoT context generation from the \textit{Brighten-Dark-Places} SmartApp.}
\label{fig:DFA}
\end{figure}

After extracting the app logic and matching the verb and noun phrases to the actual capabilities, we discover the commands from the verb phrases. For example, ``Turn on" clearly indicates the capability command \emph{capability.switch.on()}. We construct the SmartApp's behavior as an event transition graph where each node represents the capability command. The complete workflow for our example SmartApp is shown in Fig. \ref{fig:DFA}. The final event transition logic is: \emph{contactSensor.open}$\rightarrow$\emph{illuminanceMeasurement < threshold}$\rightarrow$\emph{switch.on()}, which shows the app work logic expected by user. We build the event transition graphs for all the SmartApps installed by a user, which represent the \texttt{IoT context} in the current system.



\section{Anomaly and vulnerability detection}

In this section, we introduce how to use the generated wireless context and IoT context to discover the anomalies and potential vulnerabilities in the IoT system. Each context is represented by a set of event transition graphs. We use $G= \{g_1, g_2,...,g_i,...\}$ and $G'=\{d_1,d_2,...,d_j,...\}$ to represent two sets of event transition graphs for IoT context and wireless context respectively. The nodes in the graphs $g_i$ and $d_j$ describe the corresponding IoT interaction events, which are represented by the unified capability commands. Meanwhile, all the events are numbered and given a global ID. The IoT context is the user expected app behaviors, and the wireless context is the actual app behaviors detected via the wireless communication traffic. If the wireless context violates the IoT context, that means potential anomalies and vulnerabilities. For each detected event transition graph $d_j$ in the wireless context $G'$, we check if we can find exactly the matched event transition graph $g_i$ in the IoT context $G$. The match means the $d_j$ and $g_i$ should have identical event IDs and identical event dependencies. If there is no such match, we think there is a potential anomaly in the system.

\begin{figure}[t]
\centering
\includegraphics[scale=0.15]{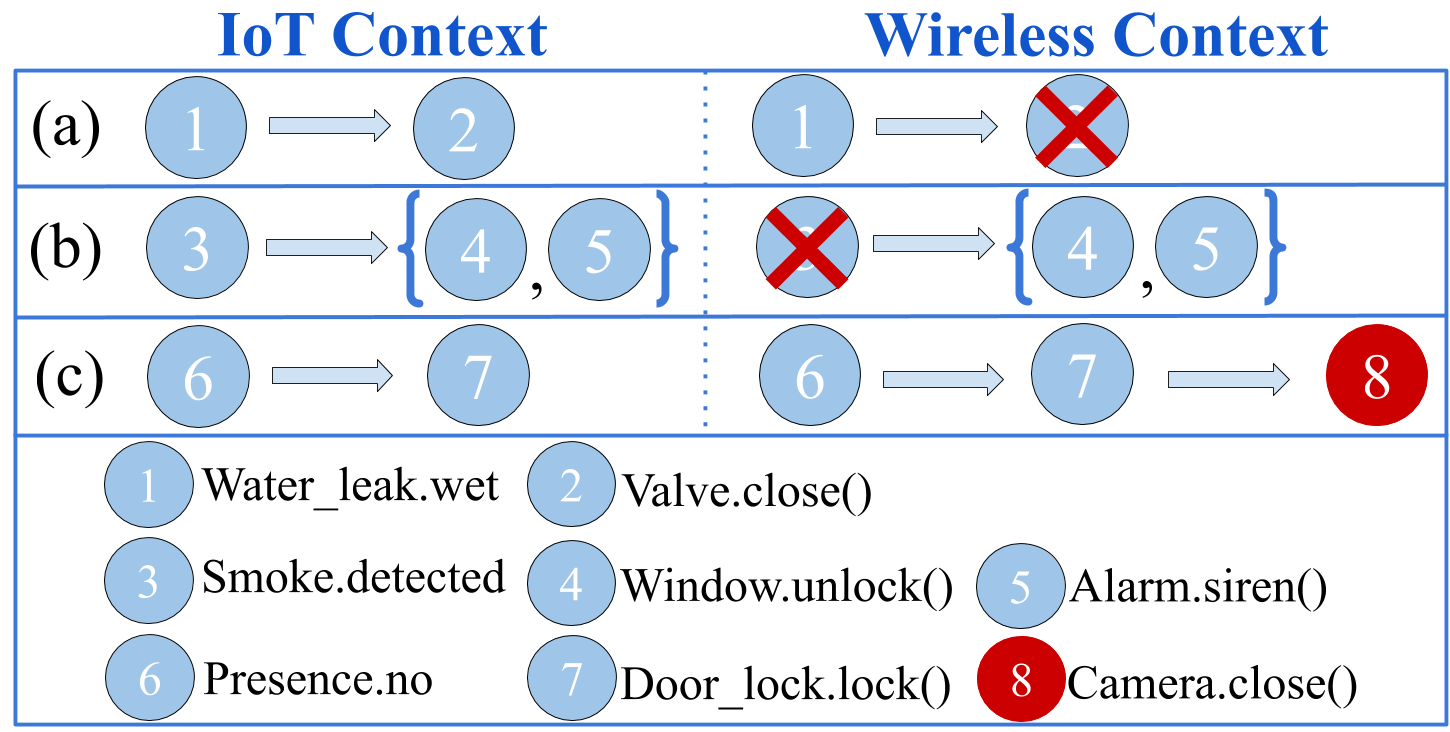}
\caption{Discovery of various anomalies.}
\label{fig:anomaly}
\end{figure}

The Fig. \ref{fig:anomaly} provide the examples of the typical anomalies we can detect via our approach. The first IoT context is ``\textit{If the water leak sensor detects the wet, close the valve}, which is represented by event transition \emph{Water\_leak.wet}$\rightarrow$\emph{Valve.close()}. For the wireless context, we only detect the first event and miss the second event. This anomaly could be caused by \texttt{device failure} or \texttt{app misbehavior}. The valve may not work due to its hardware flaws or software bugs. Also, the anomaly could be due to the app misbehavior. Once the app receives the wet alarm from the water leak sensor, it should send the command to close the valve. But from the wireless side, the app does not execute the second step. The second example is caused by \texttt{event spoofing}. Only when detecting smoke, the window is opened, and the alarm is triggered. But the attacker may spoof a fake smoke detected event and trigger subsequent actions. For the third example, we find an additional action \emph{Camera.close()} is detected due to \texttt{over-privilege}. The actual IoT context is \textit{If no presence is detected, lock the door.} The malicious app requests the non-necessary privilege for the camera and closes the camera after people leave the room. Then the attacker gets a chance to break in without the camera monitoring.

\begin{figure}[t]
\centering
\includegraphics[scale=0.15]{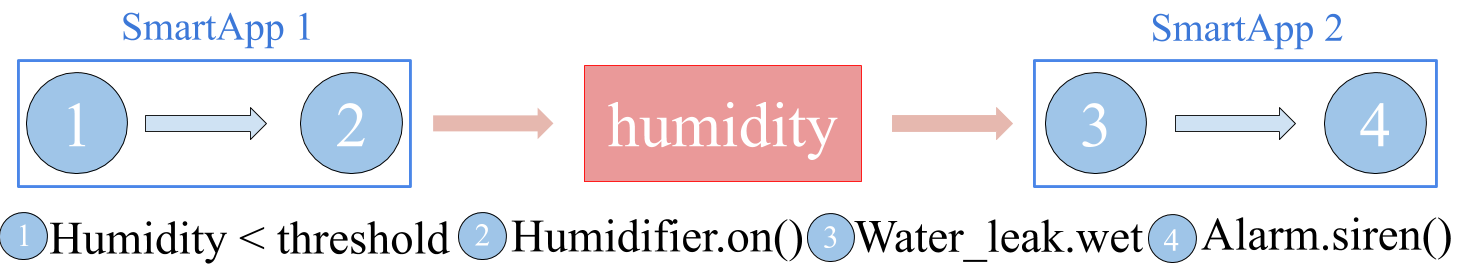}
\caption{Discovery of hidden vulnerability.}
\label{fig:vulnerability}
\end{figure}

Furthermore, our approach can detect the potential vulnerabilities that are caused by the inter-app interaction chain. Most research work focuses on the local behaviors for one single app. They ignore the potential event interaction chain that crosses multiple apps. The chain is formed via some hidden channels, such as temperature, humidity, light, etc. The formed interaction chain could be leveraged by the attacker to launch attacks. Fig. \ref{fig:vulnerability} shows an example of how to discover the vulnerabilities via our approach. For wireless context, we detect a event chain $\circled{\small 1}\xrightarrow{}\circled{\small 2}\xrightarrow{}\circled{\small 3}\xrightarrow{}\circled{\small 4}$. But we can only find the  $\circled{\small 1}\xrightarrow{}\circled{\small 2}$ and $\circled{\small 3}\xrightarrow{}\circled{\small 4}$ in the IoT context. The first app opens the humidifier once the humidity is less than a threshold. Meanwhile, the humidity could influence the input of the second app. One malicious app can change the threshold and let the humidifier keeps working until triggering the water leak alarm. \pname can detect such hidden vulnerabilities in advance and propose solutions to prevent such attacks.

\section{Evaluation}

\begin{figure}[t]
\centering
\includegraphics[scale=0.05]{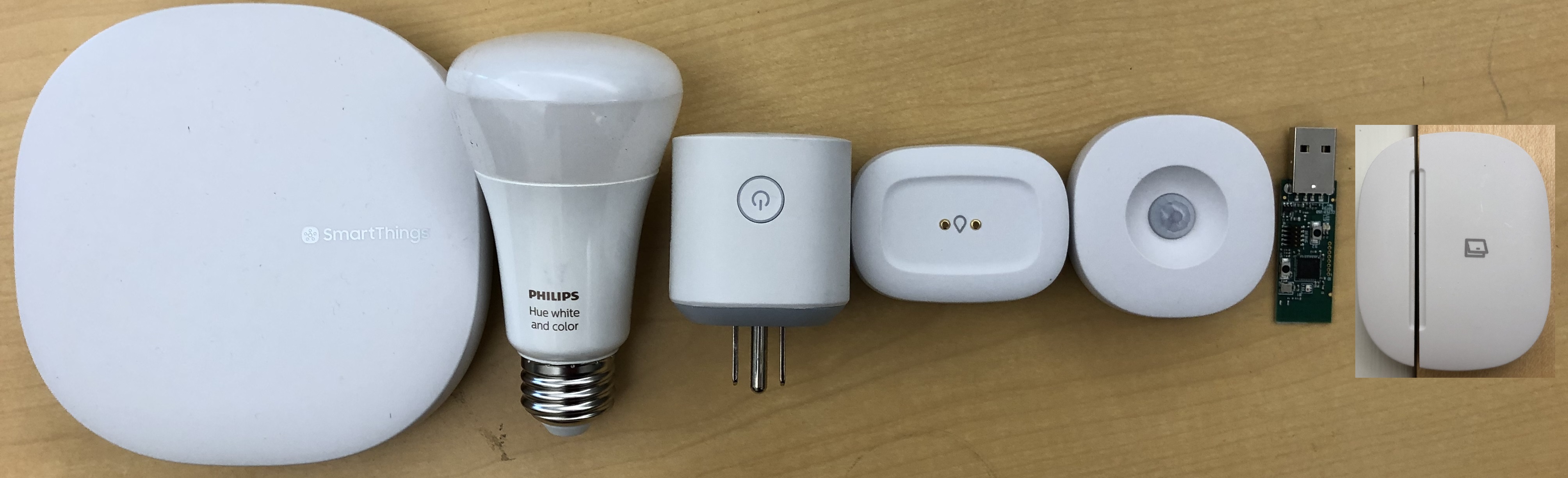}
\caption{Smart devices and ZigBee packet sniffer used in our testbed for evaluating \pname.}
\label{fig:testbed}
\end{figure}

In order to demonstrate the feasibility and effectiveness of \pname, we implement our framework on the Samsung SmartThings platform. Fig. \ref{fig:testbed} exhibits the IoT devices we use in our testbed. All the devices are connected to a SmartThings hub with ZigBee wireless communication protocol. We use TI CC2531 USB Dongle \cite{cc25312019} and install the Zigbee protocol sniffer \cite{sniffer2019} to sniff the wireless communication traffic between hub and devices. A set of SmartApps is installed to SmartThings to enable the provision of automation services. We design extensive experiments from various aspects to thoroughly evaluate our approach.

\subsection{Anomaly Detection Evaluation}
\begin{figure*}
\centering  
\subfigure[Motion sensor]{\includegraphics[height=1.21in, width=0.19\linewidth]{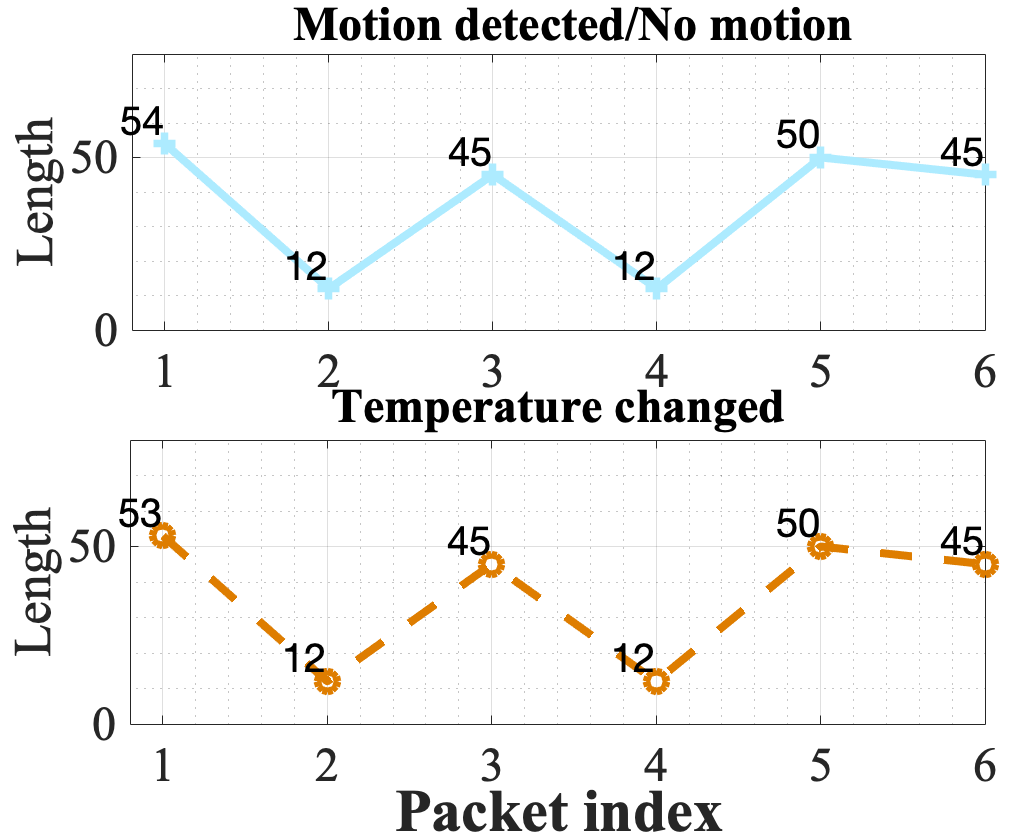}}
\subfigure[Outlet]{\includegraphics[height=1.21in, width=0.19\linewidth]{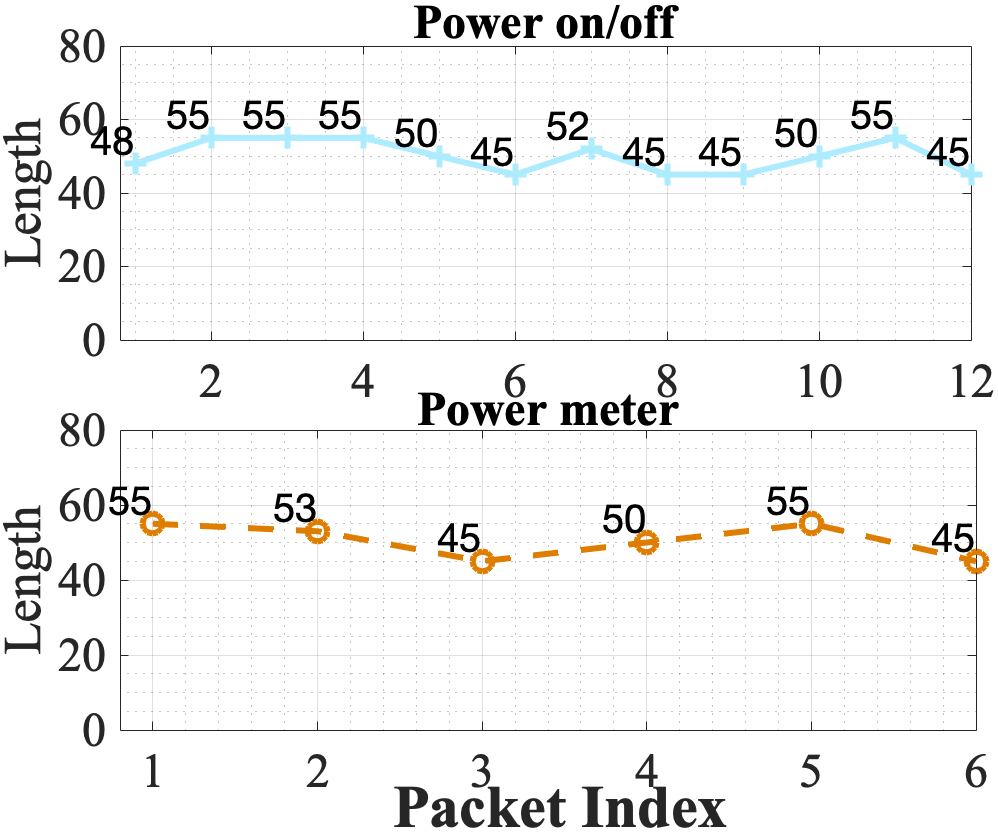}}
\subfigure[Water leak sensor]{\includegraphics[height=1.21in, width=0.19\linewidth]{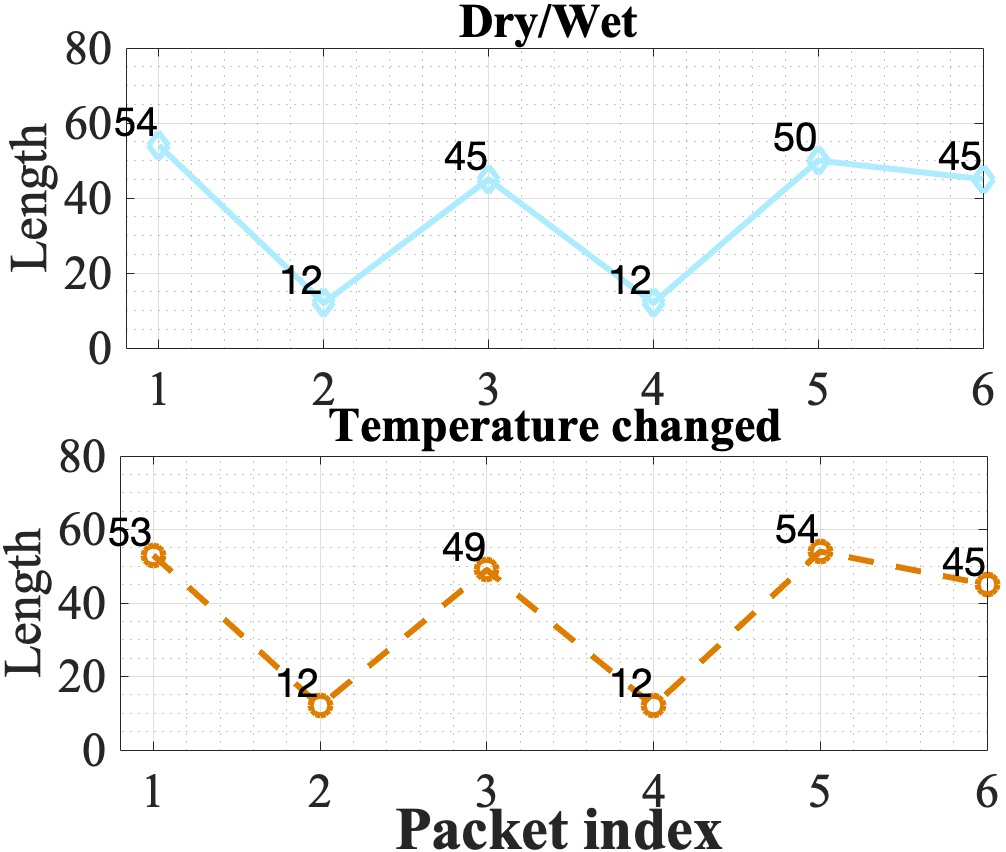}}
\subfigure[Philips Hue A19]{\includegraphics[height=1.21in, width=0.19\linewidth]{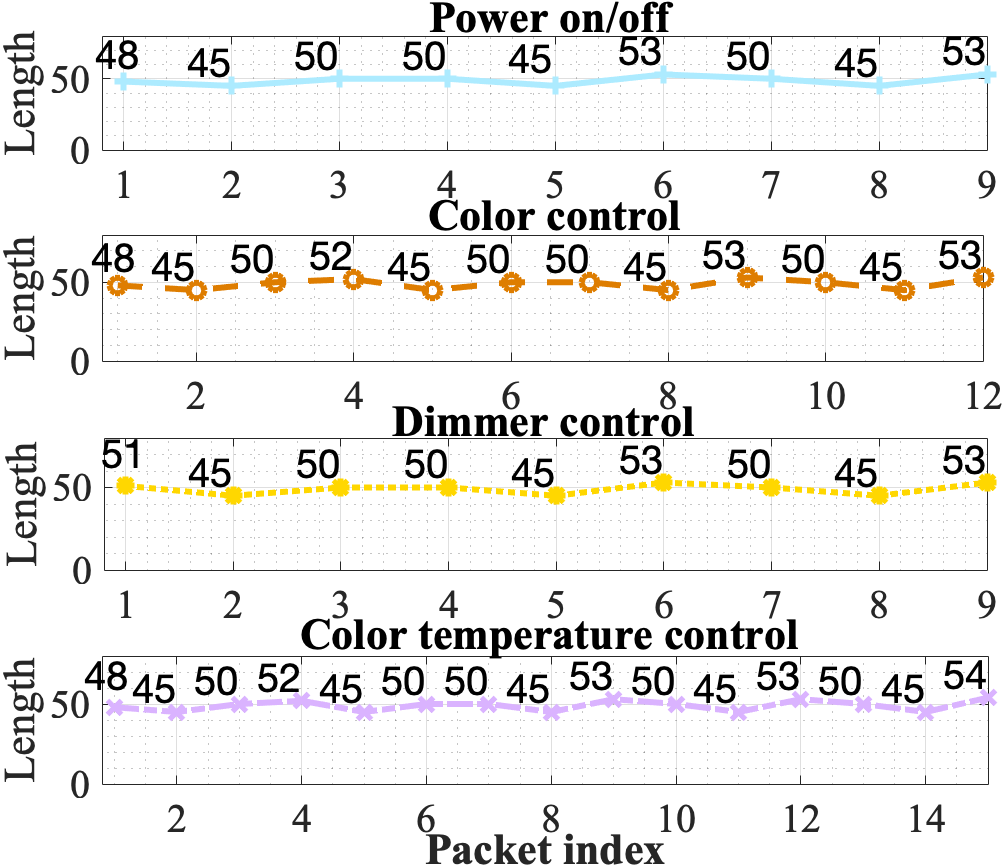}}
\subfigure[Multipurpose sensor]{\includegraphics[height=1.21in, width=0.19\linewidth]{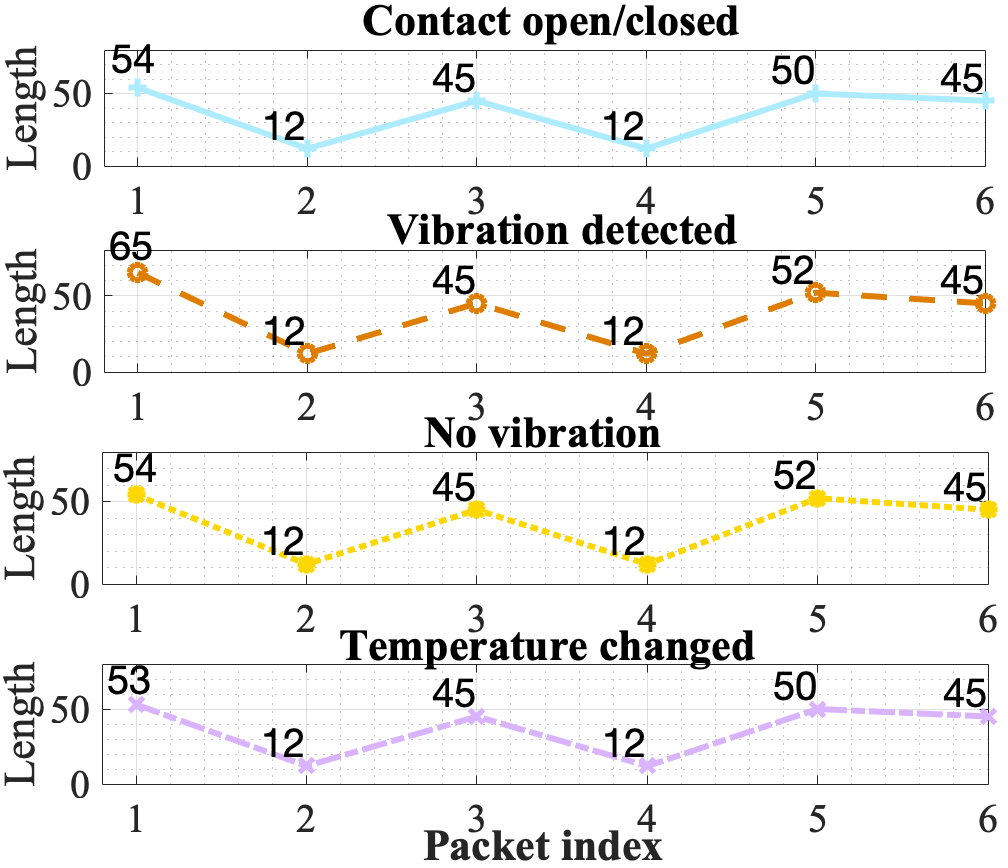}}
\caption{Packet size sequence from transmitted packets for different type events.}
\label{fig:pakcet_size_sequence}
\end{figure*}

To verify the accuracy of our event fingerprinting approach, we use the five most commonly used IoT devices for our experiments: Motion sensor, Outlet, Water leak sensor, Philips Hue A19, and Multipurpose sensor. These devices can generate 19 types of events that can be found via the SmartThings iOS app. Each event corresponds to a SmartThings capability command. For example, Fig. \ref{fig:pakcet_size_sequence}(d) shows that the Philips Hue A19 can generate the following IoT events: \textit{power on/off, color control, dimmer control,} and \textit{color temperature control.} The corresponding capability commands are \textit{switch.on(), switch.off(), colorControl.setColor(), switchLevel.setLevel(),} and \textit{colorTemperature.setColorTemperature().} Our goal is to identify these events via sniffing the wireless packets.

\begin{figure*}
\centering  
\subfigure[]{\includegraphics[height=1.21in, width=0.18\linewidth]{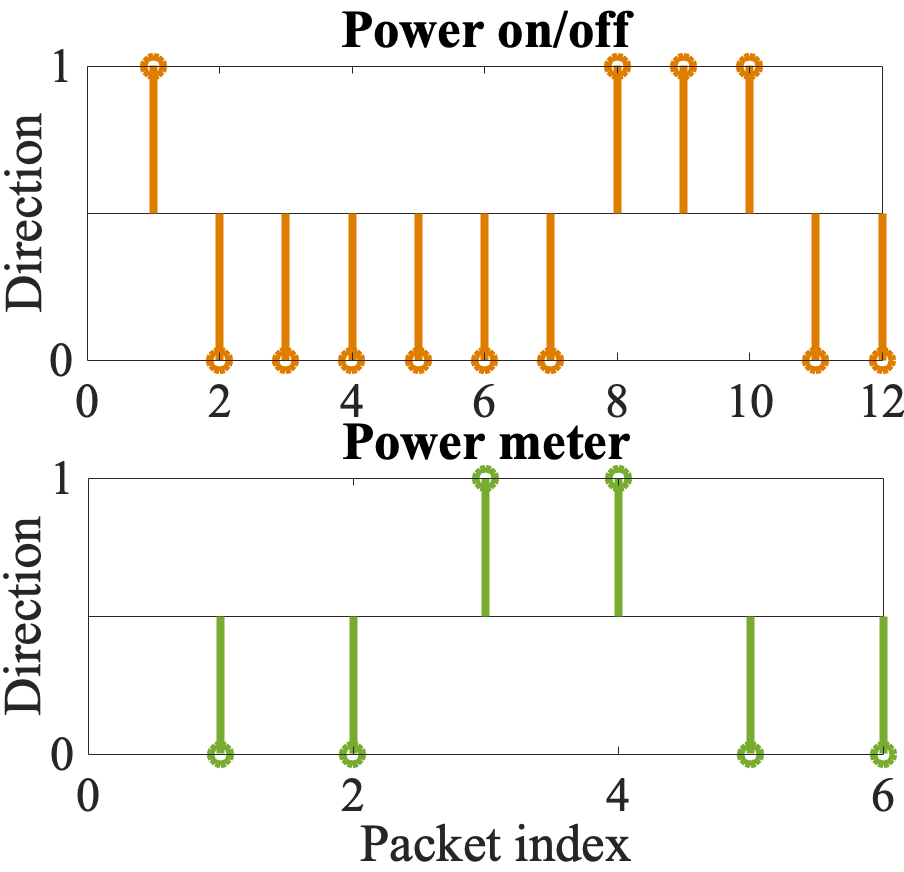}}
\subfigure[]{\includegraphics[height=1.21in, width=0.18\linewidth]{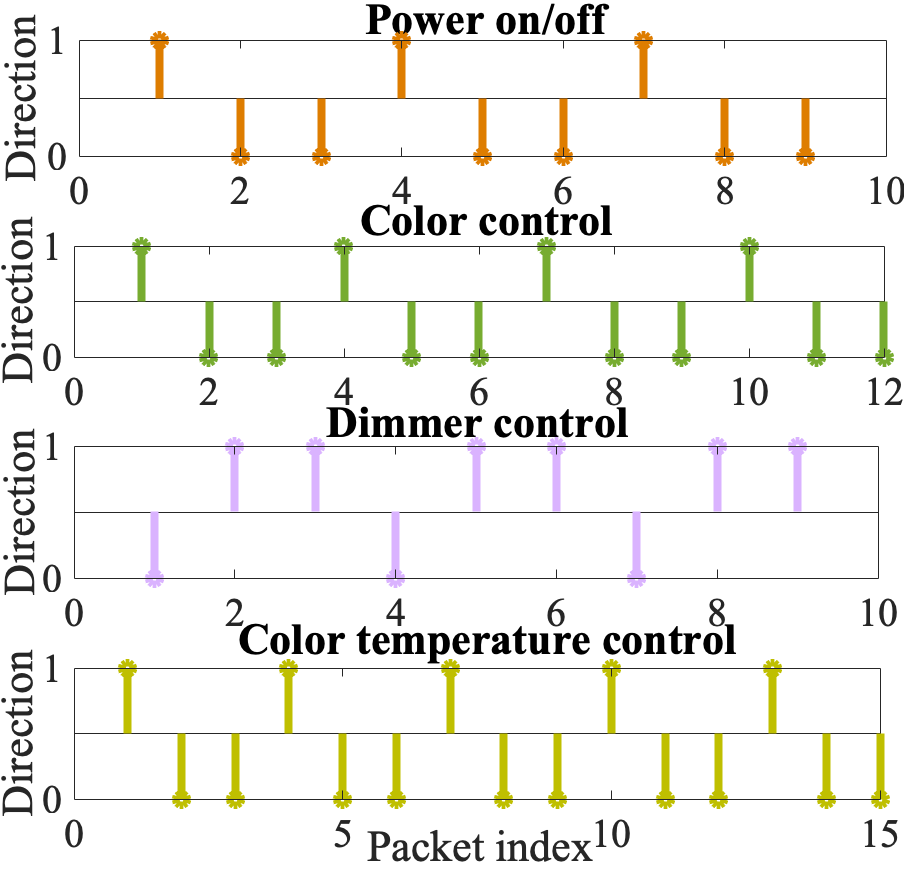}}
\subfigure[]{\includegraphics[height=1.21in, width=0.21\linewidth]{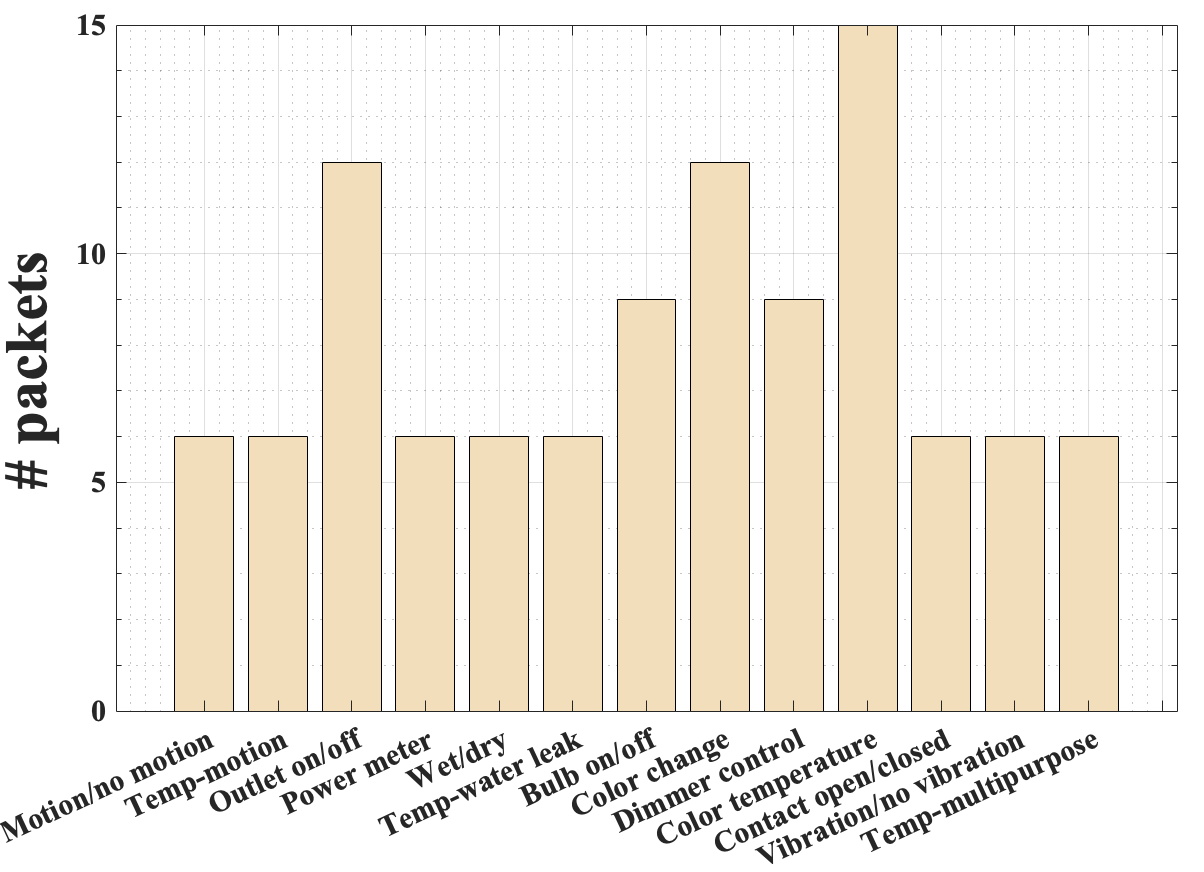}}
\subfigure[]{\includegraphics[height=1.21in, width=0.21\linewidth]{figures/bulb_length.png}}
\subfigure[]{\includegraphics[height=1.27in, width=0.19\linewidth]{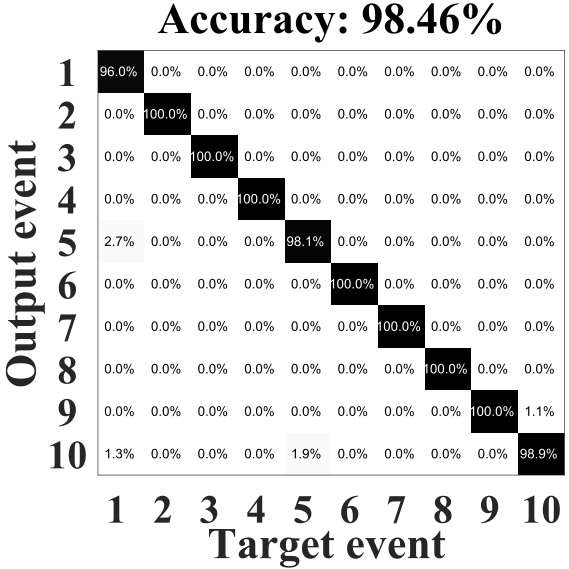}}
\caption{(a) Packet direction for outlet related events. (b) Packet direction for Philips Hue A19 related events. (c) \# packets transmission for different events. (d) Transmission time for different events. (e) Confusion matrix for event identification.}
\label{fig:wirelesseval}
\end{figure*}

\textbf{Event Fingerprinting Analysis}. We collect the sniffed Zigbee wireless traffic and correlate them with the downloaded event history from the SmartThings app. Here are observations from the experiments: (1) For most events, the packet size sequences are distinct, as shown in Fig. \ref{fig:pakcet_size_sequence}. Although the event pair \textit{motion detected/no motion}, \textit{power on/off}, and \textit{dry/wet} have the same packet length with different data fields, events in each pair are contrary to each other, which implies that we can use one variable to record the device's status to distinguish these events. (2) Although some sensors use identical capabilities for the same purpose, their packet sequence and size are still distinct. For instance, the motion sensor, water leak sensor, and multipurpose sensor can all detect temperature change. Once the temperate change is detected, they all send the same event via capability command \emph{temperature.value}. However, their packet size sequences are distinguishable and can be used for fingerprinting, as shown in Fig. \ref{fig:pakcet_size_sequence}(a)(c)(e). (3) The direction of the packet in the sequence can also be used to distinguish some events. We use \textbf{0} to denote the direction from the device to the hub and use \textbf{1} for the reverse direction. Fig. \ref{fig:wirelesseval}(a)(b) show the packet directions for different events from outlet and Philips Hue A19, and verify the effectiveness of using packet direction as a fingerprint feature. (4) The inter-packet time interval and the transmission time can also be used to distinguish different events. The transmission time for each event is shown in Fig. \ref{fig:wirelesseval}(c), from which we can see that the shortest average transmission time is 0.1477s (for power meter event), while the longest average transmission time is 2.0656s (for color change event).

\textbf{Event Collection and Model Training}. To train the event classifier $\mathcal{C}$, we deploy the testbed in a typical office environment and continuously sniff and collect three weeks' wireless packet sequence and build the fingerprint matrix for each event. Fig. \ref{fig:wirelesseval}(d) shows that the maximum number of packets transmitted, for all kinds of events, is 15. So we set the parameter of $\mathcal{N}$ defined in section \ref{sec:fingerprint} to be 15. We label the data by matching the recorded event history in the SmartThings app. For devices that are triggered very infrequently in a real office environment, such as water leak sensor, we manually wet and dry it to generate sufficient event samples for the training. We train a random forest classifier using the event samples.

\textbf{Event Detection Analysis}. Based on the devices' capabilities in our testbed, we select the existing apps in the SmartThings Public Github Repository \cite{appslist2019} and also develop our customized apps to build a total of 35 apps library and install them in the SmartThings platform. We constantly sniff the wireless traffic for another one week and identify the event in a real-time manner and generate the event sequence. We set the value of parameter $\mathcal{T}$ and $\theta$ in Section \ref{sec:event_identifcation} to be 2.1s and 0.7 respectively. Before the identification, we remove the unrelated packets, including beacon packets, link-maintain packets, and acknowledgment packets. We still compare the detected events with the recorded events in the SmartThings hub and compute the detection accuracy. We select the ten most frequently occurred events and show their confusion matrix for classification in Fig. \ref{fig:wirelesseval}(e). The overall identification accuracy is 98.46\%. The detection failure is mainly due to packet loss --- the sniffer may miss some packets due to its limitation or other signal interference.

\begin{table*}[t]
\centering
\caption{Anomaly detection via the discovery and comparison of IoT and wireless context.}
\begin{tabular}[c]{|M{0.4cm}|M{12.6cm}|M{3.4cm}|}
\hline
\rowcolor{Gray}
\textbf{No.} & \textbf{Event dependencies discovered in IoT context}             & \textbf{Detected wireless context} \\ \hline
1
& \textcolor{black}{\framebox[1.1\width]{motion sensor}}\textcolor{black}{$\xrightarrow{\text{motion.active}}$} \framebox[1.1\width]{hub} $\xrightarrow{\text{switch.on()}}$ \framebox[1.1\width]{\smash{P}hili\smash{p}s Hue} 
&  \textcolor{red}{\sout{\circled{\small 1}}} $\xrightarrow{} \circled{\small 2}$  
\\ \hline
2
& \textcolor{black}{\framebox[1.1\width]{multi\smash{p}ur\smash{p}ose sensor}}\textcolor{black}{$\xrightarrow{\text{temperature.value}}$} \framebox[1.1\width]{hub} $\xrightarrow{\text{colorControl.setColor()}}$ \framebox[1.1\width]{\smash{P}hili\smash{p}s Hue}
&  \textcolor{red}{\sout{\circled{\small 3}}} $\xrightarrow{} \circled{\small 4}$ 
\\ \hline
3
& \textcolor{black}{\framebox[1.1\width]{outlet}}\textcolor{black}{$\xrightarrow{\text{power.value}}$} \framebox[1.1\width]{hub} $\xrightarrow{\text{switch.off()}}$ \framebox[1.1\width]{outlet}  
&  \textcolor{red}{\sout{\circled{\small 5}}} $\xrightarrow{} \circled{\small 6}$  
\\ \hline
4   
& \textcolor{black}{\framebox[1.1\width]{water leak sensor}}\textcolor{black}{$\xrightarrow{\text{water.wet}}$} \framebox[1.1\width]{hub} $\xrightarrow{\text{switch.off()}}$ \framebox[1.1\width]{outlet}
&  \circled{\small 7} $\xrightarrow{}$ \textcolor{red}{\sout{\circled{\small 6}}}
\\ \hline
5   
& \textcolor{black}{\framebox[1.1\width]{multi\smash{p}ur\smash{p}ose sensor}}\textcolor{black}{$\xrightarrow{\text{acceleration.active}}$} \framebox[1.1\width]{hub} $\xrightarrow{\text{switch.on()}}$ \framebox[1.1\width]{\smash{P}hili\smash{p}s Hue}
& \circled{\small 8} $\xrightarrow{}$ \textcolor{red}{\sout{\circled{\small 2}}} 
\\ \hline
6   
& \textcolor{black}{\framebox[1.1\width]{multi\smash{p}ur\smash{p}ose sensor}}\textcolor{black}{$\xrightarrow{\text{contact.open}}$} \framebox[1.1\width]{hub} $\xrightarrow{\text{switch.on()}}$ \framebox[1.1\width]{\smash{P}hili\smash{p}s Hue}
& \circled{\small 9} $\xrightarrow{}$ \textcolor{red}{\sout{\circled{\small 2}}} 
\\ \hline
7   
& \textcolor{black}{\framebox[1.1\width]{multi\smash{p}ur\smash{p}ose sensor}}\textcolor{black}{$\xrightarrow{\text{contact.close}}$} \framebox[1.1\width]{hub} $\xrightarrow{\text{switch.off()}}$ \framebox[1.1\width]{outlet} 
& \circled{\small 9} $\xrightarrow{}$ \circled{\small 6} \textcolor{red}{$\xrightarrow{}$ \circled{\small 2}}
\\ \hline
8  
& \textcolor{black}{\framebox[1.1\width]{motion sensor}}\textcolor{black}{$\xrightarrow{\text{motion.inactive}}$} \framebox[1.1\width]{hub} $\xrightarrow{\text{colorControl.setHue()}}$ \framebox[1.1\width]{\smash{P}hili\smash{p}s Hue}
& \circled{\small 1} $\xrightarrow{}$ \circled{\scriptsize 10} \textcolor{red}{$\xrightarrow{}$ \circled{\small 2}}
\\ \hline
9
& \framebox[1.1\width]{hub} $\xrightarrow{\text{switch.off()}}$ \framebox[1.1\width]{bulb}, \framebox[1.1\width]{hub} $\xrightarrow{\text{lock.lock()}}$ \framebox[1.1\width]{lock}, \framebox[1.1\width]{hub} $\xrightarrow{\text{switch.on()}}$ \framebox[1.1\width]{camera}
& \circled{\scriptsize 11} $\xrightarrow{}$ \circled{\scriptsize 12} $\xrightarrow{}$ \textcolor{red}{\sout{\circled{\scriptsize 13}}} 
\\ \hline
10
& $\bm{\{}$ \framebox[1.1\width]{multi\smash{p}ur\smash{p}ose} $\xrightarrow{\text{contact.open}}$, \framebox[1.1\width]{illuminance sensor} $\xrightarrow{\text{illuminance.value}}$ $\bm{\}}$ \framebox[1.1\width]{hub} $\xrightarrow{\text{switch.on()}}$ \framebox[1.1\width]{bulb}
& $\bm{\{}$ \textcolor{red}{\sout{\circled{\small 9}}}, \circled{\scriptsize 14} $\bm{\}}$ $\xrightarrow{}$ \circled{\scriptsize 11}
\\ \hline
11
& \framebox[1.1\width]{multi\smash{p}ur\smash{p}ose} $\xrightarrow{\text{temperature.value}}$  \framebox[1.1\width]{hub} 
$\bm{\{}$ $\xrightarrow{\text{colorControl.setColor()}}$ \framebox[1.1\width]{Hue}, $\xrightarrow{\text{switch.on()}}$ \framebox[1.1\width]{heater} $\bm{\}}$
& \circled{\small 3} $\xrightarrow{}$ $\bm{\{}$ \textcolor{red}{\sout{\circled{\small 4}}}, \textcolor{red}{\sout{\circled{\scriptsize 15}}} $\bm{\}}$
\\ \hline
12
& $\bm{\{}$ \framebox[1.1\width]{thermostat} $\xrightarrow{\text{presence.not\_present}}$, \framebox[1.1\width]{multi\smash{p}ur\smash{p}ose} $\xrightarrow{\text{contact.closed}}$, \framebox[1.1\width]{lock}$\xrightarrow{\text{lock.unlocked}}$ $\bm{\}}$ \framebox[1.1\width]{hub} $\xrightarrow{\text{lock.lock()}}$ \framebox[1.1\width]{lock}
& $\bm{\{}$ \circled{\scriptsize 16}, \circled{\scriptsize 17}, \circled{\scriptsize 18} $\bm{\}}$ $\xrightarrow{}$ \textcolor{red}{\sout{\circled{\scriptsize 12}}}
\\ \hline
\end{tabular}
\label{table:context_generation}
\end{table*}

\textbf{Wireless and IoT Context Discovery}. After generating the sequence of the events, we mine the event dependencies using our algorithm and discover the wireless context. We successfully detect wireless context consisting of 35 event dependencies. The first eight items in Table \ref{table:context_generation} show part of the detected wireless context. We also use the proposed NLP approach to extract the IoT context from 35 apps installed, which exactly matches the detected wireless context. 
To further evaluate the applicability of our approach, we generate some complicated event dependencies, shown in the last four items in Table  \ref{table:context_generation}. We insert these events to the sequence of the already existing events and verify that we can still discover these complicated wireless context. The experimental results demonstrate the effectiveness of our IoT context and wireless context discovery.

\textbf{Anomaly Generation and Detection}. For the installed 35 apps, we design and insert malicious code to the apps to generate anomalies. For each app, we modify its code to generate the following three types of anomalies: (a) Event Spoofing. We add the code in the app to spoof some events for triggering purpose. The first three items in Table \ref{table:context_generation} show the examples, such as event sequence changing from $\circled{\small 1}\xrightarrow{}\circled{\small 2}$ to \textcolor{red}{\sout{\circled{\small 1}}}$\xrightarrow{}\circled{\small 2}$. (b) App Misbehavior. For the item 4-6 in Table \ref{table:context_generation}, we modify the apps' code and make the app \emph{not} execute the triggered actions, leading to the event sequence change from $\circled{\small 7}\xrightarrow{}\circled{\small 6}$ to $\circled{\small 7}\xrightarrow{}$ \textcolor{red}{\sout{\circled{\small 6}}}. (c) Over-privilege. The item 7-8 show the anomaly samples we generate. We modify the code to request the non-necessary capabilities and execute the addition actions, such as changing from $\circled{\small 9}\xrightarrow{}\circled{\small 6}$ to $\circled{\small 9}\xrightarrow{}\circled{\small 6}\enspace\textcolor{red}{\xrightarrow{}\circled{\small 2}}$. For each app, we generate 100 anomalies for each threat type and try to detect them via our approach.

The results of anomaly detection are shown in Table \ref{table:detection}. We can see that the detection for overprivilege has both high precision and recall. This is because the overprivileged event sequence is different from all of the normal sequences, so the precision and recall are just limited by the success rate of event detection. Spoofing and misbehavior can occasionally generate event sequence which match the normal app behavior. Therefore, their precision is high but recall is low.

\begin{table}[t]
\centering
\caption{Anomaly detection results for three types of anomaly.}
\begin{tabular}{|>{\columncolor[gray]{0.9}}c|c|c|c|}
\hline
\rowcolor{Gray}
          & \textbf{Spoofing} & \textbf{Overprivilege} & \textbf{Misbehavior} \\ \hline
\textbf{Precision} & 97.22\%  & 98.55\%       & 98.29\%     \\ \hline
\textbf{Recall}    & 94.82\%  & 98.36\%       & 95.20\%     \\ \hline
\end{tabular}
\label{table:detection}
\end{table}

\begin{table*}[t]
\centering
\caption{Hidden vulnerabilities discovery via analyzing wireless context.}
\begin{tabular}[c]{|M{0.4cm}|M{16cm}|}
\hline
\rowcolor{Gray}
\textbf{No.} & \textbf{Event dependencies discovered in wireless context}  \\ \hline
1
& {\framebox[1.1\width]{time}$\xrightarrow{}$\framebox[1.1\width]{hub}$\xrightarrow{\text{switch.on()}}$ \framebox[1.1\width]{heater}} $\xrightarrow{}$ \textcolor{orange}{\framebox[1.1\width]{\textbf{temperature}}} $\xrightarrow{}$ {\framebox[1.1\width]{temperature sensor} $\xrightarrow{\text{temperature.value}}$ \framebox[1.1\width]{hub} $\xrightarrow{\text{window.open()}}$ \framebox[1.1\width]{window}}
\\ \hline
2
& {\framebox[1.1\width]{tem\smash{p}erature sensor} $\xrightarrow{\text{temperature.value}}$ \framebox[1.1\width]{hub}$\xrightarrow{\text{switch.on()}}$ \framebox[1.1\width]{fan}} $\xrightarrow{}$ 
\textcolor{orange}{\framebox[1.1\width]{\textbf{motion}}}
$\xrightarrow{}$ {\framebox[1.1\width]{motion sensor} $\xrightarrow{\text{motion.active}}$ \framebox[1.1\width]{hub}$\xrightarrow{\text{switch.on()}}$ \framebox[1.1\width]{\smash{l}i\smash{g}ht}} 
\\ \hline
3
& {\framebox[1.1\width]{water leak sensor} $\xrightarrow{\text{water.wet}}$ \framebox[1.1\width]{hub}$\xrightarrow{\text{switch.on()}}$ \framebox[1.1\width]{light}} $\xrightarrow{}$ \textcolor{orange}{\framebox[1.1\width]{\textbf{illuminance}}}
$\xrightarrow{}$ {\framebox[1.1\width]{illum sensor} $\xrightarrow{\text{illuminance.value}}$ \framebox[1.1\width]{hub}$\xrightarrow{\text{windowShade.close()}}$ \framebox[1.1\width]{window shade}} 
\\ \hline
4
& {\framebox[1.1\width]{presence sensor} $\xrightarrow{\text{presence.not\_present}}$ \framebox[1.1\width]{hub}
$\xrightarrow{\text{lock.lock()}}$}
\textcolor{orange}{\framebox[1.1\width]{\textbf{lock}}}
{$\xrightarrow{\text{lock.lock}}$ 
\framebox[1.1\width]{hub}
$\xrightarrow{\text{colorControl.setColor()}}$ \framebox[1.1\width]{Hue}} 
\\ \hline
5
& {\framebox[1.1\width]{presence sensor} $\xrightarrow{\text{presence.present}}$ \framebox[1.1\width]{hub}
$\xrightarrow{\text{switch.on()}}$}
\textcolor{orange}{\framebox[1.1\width]{\textbf{bulb}}}
{$\xrightarrow{\text{switch.on}}$ 
\framebox[1.1\width]{hub}
$\xrightarrow{\text{camera.take()}}$ \framebox[1.1\width]{camera}} 
\\ \hline
6
& {\framebox[1.1\width]{multi\smash{p}ur\smash{p}ose sensor} $\xrightarrow{\text{temperature.value}}$ \framebox[1.1\width]{hub}
$\xrightarrow{\text{switch.on()}}$}
\textcolor{orange}{\framebox[1.1\width]{\textbf{AC}}}
{$\xrightarrow{\text{switch.on}}$ 
\framebox[1.1\width]{hub}
$\xrightarrow{\text{switch.on()}}$ \framebox[1.1\width]{bulb}} 
\\ \hline
7
& {\framebox[1.1\width]{\smash{p}resence.not\_\smash{p}resent}
$\xrightarrow{\text{presence.not\_present}}$ 
\framebox[1.1\width]{hub}
$\xrightarrow{}$}
\textcolor{orange}{\framebox[1.1\width]{\textbf{location mode}}}
{$\xrightarrow{}$
\framebox[1.1\width]{hub} 
$\xrightarrow{\text{switch.off()}}$ 
\framebox[1.1\width]{\smash{l}i\smash{g}ht}}
\\ \hline
\end{tabular}
\label{table:physicalchannel}
\end{table*}

\begin{table}[t]
\centering
\caption{Statistics of hidden channels identified from official SmartApps.}
\begin{tabular}[c]{|M{1.2cm}|M{2cm}|M{1.5cm}|M{1.6cm}|}
\hline
\rowcolor{Gray}
\textbf{Channel Type} & \textbf{Channel} & \textbf{\# apps related} & \textbf{\# interaction chains} \\ \hline
\multicolumn{1}{|c|}{\multirow{7}{*}{\shortstack[l]{Capability}}} & swtich(light)  & 28            & 127                    \\ \cline{2-4} 
\multicolumn{1}{|c|}{}                                    & doorControl    & 4             & 4                      \\ \cline{2-4} 
\multicolumn{1}{|c|}{}                                    & lock           & 8             & 22                     \\ \cline{2-4} 
\multicolumn{1}{|c|}{}                                    & switch(heater) & 10            & 27                     \\ \cline{2-4} 
\multicolumn{1}{|c|}{}                                    & switch(AC)     & 9             & 23                     \\ \cline{2-4} 
\multicolumn{1}{|c|}{}                                    & colorControl   & 7             & 6                      \\ \cline{2-4} 
\multicolumn{1}{|c|}{}                                    & thermostat     & 9             & 20                     \\ \hline
\multirow{9}{*}{\shortstack[l]{Physical}}                         & leakage        & 4             & 5                      \\ \cline{2-4} 
                                                          & illuminance    & 29            & 132                    \\ \cline{2-4} 
                                                          & energy         & 36            & 134                    \\ \cline{2-4} 
                                                          & contact        & 20            & 37                     \\ \cline{2-4} 
                                                          & acceleration   & 9             & 18                     \\ \cline{2-4} 
                                                          & smoke          & 10            & 17                     \\ \cline{2-4} 
                                                          & temperature    & 18            & 127                    \\ \cline{2-4} 
                                                          & motion         & 14            & 13                     \\ \cline{2-4} 
                                                          & humidity       & 4             & 3                      \\ \hline
\multirow{1}{*}{\shortstack[l]{System}}                                            & location.mode  & 9             & 16                     \\ \hline
\end{tabular}
\label{tab:channel}
\end{table}

\subsection{Hidden Vulnerabilities Discovery}
The current research mostly focuses on security vulnerability detection per SmartApp. The local behaviors of one single app may explicitly or implicitly affect the whole IoT system. The potential interactions between apps, devices, and the environment may produce vulnerabilities that cannot be discovered by per-app analysis. We propose to use the wireless context to discover the hidden vulnerabilities that also can be exploited by attackers. 

In wireless context, we can find some wireless event dependencies spanning multiple applications. This is because these applications are somehow \emph{correlated} together via some hidden channels. We thoroughly investigate the 183 apps in the SmartThings Public GitHub Repository \cite{appslist2019} and analyze their interactions with other apps, devices, and the environment. We find that there are three kinds of channels that can cause potential vulnerabilities: (1) \textbf{Capability}. Two applications can interact if the first app's output is the trigger of the second app. For example, the SmartApp ``NFC Tag Toggle'' in the official SmartThings GitHub allows toggling of a switch, lock, or garage door. And another SmartApp ``Door State to Color Light (Hue Bulb)'' changes the color of Hue bulbs based on the door status. In this example, the two apps are directly chained via the capability \emph{doorControl}. (2) \textbf{Physical channel}. The environment elements can be changed due to the input or output of some apps and cause potential interaction chains. We take one physical channel \textit{smoke} as an example. The toaster may cause smoke, which makes alarm siren. (3) \textbf{System channel}. Some global variables in the IoT program framework may be shared by some SmartApps. The \emph{location.mode} in SmartThings platform enables the devices to behave differently in different scenarios. For example, if the current location.mode is ``Home'' and the motion sensor detects motion, then turn on the light. But if the location.mode is ``Away'' and the motion is detected, then turn on the camera. All these three kinds of channels can generate unexpected application interaction, rendering system vulnerabilities.

We discover and provide the list of all the hidden vulnerabilities for each type of channel from the SmartThings platform. Based on the NLP approach in Section \ref{sec:context}, the capabilities related to the apps' input and output are extracted. We list seven capabilities that are shared by apps in Table \ref{tab:channel}. The capability \emph{switch(light)} generates 127 potential inter-app interaction chains and has the highest risk score. We use Word2Vec \cite{mikolov2013efficient} to establish the mapping between physical channels and apps' input and output. In total, nine physical channels are discovered, as shown in Table \ref{tab:channel}. The illuminance, energy, and temperature are the channels that bring the most of inter-app interactions. When we program the malicious apps, we show that system variable \emph{location.mode} from SmartThings program platform \emph{location.mode} are frequently used and modified by some apps, which can also cause security-relevant issues. We list the vulnerabilities found for each type of channel in Table \ref{table:physicalchannel} and show the statistics about vulnerabilities in Table \ref{tab:channel}. 


\section{Related Work}

IoT system is composed of protocols, devices, apps, platforms, and the environment. The complexity of the IoT system makes it challenging to resolve security and privacy issues. Each component in the IoT system can cause potential threats\cite{fang2019foresee}. The device flaws \cite{sivaraman2016smart,ronen2017iot,ronen2017iot} can be exploited by attackers to infiltrate the IoT networks. For smart apps, the static and dynamic program analysis \cite{jia2017contexlot,celik2018sensitive,celik2018soteria,celik2019iotguard} are used to track the apps' control and data flow so as to prevent the sensitive data leakage and identify the potential app misbehavior. The research work \cite{fernandes2016security,fernandes2016flowfence,wang2018fear} focus on the platform security and try to exploit the design flaws of exiting program frameworks and propose solutions to prevent app over privilege and sensitive information leakage. For instance, 
The authors \cite{wang2018fear} propose to collect provenance of events and data state changes to build provenance graphs of their causal relationships, enabling attack detection.

Some other techniques are also used to enhance the IoT security. The device fingerprinting technique is developed in \cite{han2018you,formby2016s} to distinguish between legitimate devices and attacker devices. By analyzing the encrypted network traffic, \cite{taylor2016appscanner} can build app fingerprints and \cite{miettinen2017iot,gu2018bf} can build the fingerprints for identifying the types of devices. Model checking is used in \cite{nguyen2018iotsan} as a building block to reveal ``interaction-level'' flaws by identifying events that can lead the system to unsafe states. Graph-based detection approaches \cite{gcnjiuming, wanderjiuming} can also be applied to IoT to detect anomalies. Natural language processing (NLP) is used in mobile apps \cite{pandita2013whyper, pan2018flowcog, zhang2015towards} and IoT apps \cite{tian2017smartauth, ding2018safety} to automatically extract security-relevant information from apps' description, code, and annotations. The extracted semantics are compared to the tracked control and data flows in the program so as to detect apps' misbehaviors, which require complicated program analysis techniques. Our approach considers the anomaly detection starting from the view of wireless context. HoMonit\cite{zhang2018homonit} has a similar idea with us, and it compares the IoT activities inferred from the encrypted traffic with their expected behaviors dictated in their source code. But, our work does not need to analyze the source code and mainly focuses on the discovery of wireless context. We generate the sequential IoT events and mine their temporal event dependencies to explore all actual wireless context, and then compare with the IoT context inferred from apps' descriptions. By analyzing the wireless context, we can also provide a new approach to discover the hidden vulnerabilities, which HoMonit does not support.

\section{Conclusion}

In this paper, we propose a novel IoT anomaly detection framework called \pname. Instead of exploring the threats inside platform and apps, we deploy a third-party monitor \pname, who gazes at the wireless traffic and detects the potential threats in the IoT system via analyzing the encrypted wireless packets. We propose a new concept called \texttt{wireless context} in IoT that represents the observed app logic from wireless sniffing. We design a fingerprinting based event detection approach and use it to generate the event sequence via sniffed wireless packets. We design an algorithm to discover the temporal event dependencies and build the wireless context. We also extract the IoT context that reflects user expected app behaviors via analyzing apps' descriptions via natural language processing techniques. By matching the wireless and IoT context, we can detect the anomalies that are happening in the IoT system. Furthermore, the event dependencies discovered by \pname can reveal some potential vulnerabilities that are caused by the inter-app interaction via some hidden channels. We prototype our approach on the Samsung SmartThings platform and demonstrate the feasibility and effectiveness of \pname.

\bibliographystyle{ieeetr}
\bibliography{references}

\end{document}